\documentclass[9pt,twocolumn,twoside]{extarticle}
\usepackage[letterpaper,left=0.72in,right=0.72in,top=0.72in,bottom=0.70in,columnsep=0.25in]{geometry}
\usepackage[T1]{fontenc}
\usepackage{newtxtext,newtxmath}
\usepackage{graphicx,amsmath,booktabs,xcolor,cite}
\usepackage[font=small,labelfont=bf,labelsep=period]{caption}
\usepackage{titlesec,fancyhdr,stfloats,needspace}
\usepackage{flushend}
\usepackage[colorlinks=true,linkcolor=black,citecolor=black,urlcolor=black]{hyperref}
\titleformat{\section}{\sffamily\bfseries\small}{\thesection.}{0.5em}{\MakeUppercase}
\titlespacing*{\section}{0pt}{9pt}{5pt}

\newcommand{\Th}{\ensuremath{^{229}\mathrm{Th}}}

\newenvironment{acknowledgments}{\section*{Acknowledgments}}{}
\begin{document}
\twocolumn[\begin{@twocolumnfalse}
\begin{center}
{\sffamily\bfseries\fontsize{20}{23}\selectfont Continuous-wave 148.4-nm generation in strontium tetraborate for nuclear absorption spectroscopy\par}
\vspace{6pt}
{\sffamily\fontsize{10}{13}\selectfont
Hanlin~Wang\textsuperscript{1,2,\textdagger},
Zhouzhi~Tan\textsuperscript{1,2,\textdagger},
Conghui~Huang\textsuperscript{5},
Jinming~Liu\textsuperscript{2,3},
Xuan~Fan\textsuperscript{2,3},
Runzhe~Yang\textsuperscript{4},
Meng~Wang\textsuperscript{2,3},
Tung-Hsun~Chung\textsuperscript{2},
Yong-Heng~Huo\textsuperscript{2,3,8},
Yin~Hang\textsuperscript{4},
Zhiwei~Jiao\textsuperscript{6},
Heng~Tu\textsuperscript{7},
Lei~Zhang\textsuperscript{5},
Xu-Fei~Yin\textsuperscript{3,9},
Pengfei~Wang\textsuperscript{2,10},
and Zhiqiang~Zhang\textsuperscript{1,2,3,11}\par}
\vspace{5pt}
{\fontsize{8}{10}\selectfont
\textsuperscript{1}School of Future Technology, University of Science and Technology of China, Hefei 230026, China\par
\textsuperscript{2}Hefei National Laboratory, University of Science and Technology of China, Hefei 230088, China\par
\textsuperscript{3}Hefei National Research Center for Physical Sciences at the Microscale and School of Physical Sciences, University of Science and Technology of China, Hefei 230026, China\par
\textsuperscript{4}Shanghai Institute of Optics and Fine Mechanics, Chinese Academy of Sciences, Shanghai 201800, China\par
\textsuperscript{5}Shanghai Precilasers Technology Co., Ltd., Shanghai 201808, China\par
\textsuperscript{6}School of Mechanical and Materials Engineering, North China University of Technology (NCUT), Beijing 100144, China\par
\textsuperscript{7}Technical Institute of Physics and Chemistry, Chinese Academy of Sciences, Beijing 100190, China\par
\textsuperscript{8}Shanghai Research Center for Quantum Science and CAS Center for Excellence in Quantum Information and Quantum Physics, University of Science and Technology of China, Shanghai 201315, China\par
\vspace{3pt}
\textsuperscript{\textdagger}These authors contributed equally to this work.\par
Co-corresponding authors: Xu-Fei Yin, Pengfei Wang, and Zhiqiang Zhang.\par
\textsuperscript{9}e-mail: \href{mailto:xufeiyin@ustc.edu.cn}{xufeiyin@ustc.edu.cn}\quad
\textsuperscript{10}e-mail: \href{mailto:wangpf@hfnl.cn}{wangpf@hfnl.cn}\quad
\textsuperscript{11}e-mail: \href{mailto:zhang-zhiqiang@ustc.edu.cn}{zhang-zhiqiang@ustc.edu.cn}\par
}
\vspace{6pt}
\begin{minipage}{0.96\textwidth}
\textbf{Abstract.} Continuous-wave light near 148.4~nm enables direct absorption spectroscopy of the thorium-229 nuclear transition. We demonstrate single-pass frequency doubling in strontium tetraborate, delivering an estimated 8.1~nW at 920~mW pump power. At this power, oxygen admission leaves a background of 1.22~kHz, corresponding to an estimated signal-to-background ratio of $3.8\times10^4$ after current-integration calibration. Several crystals grown by top-seeded solution growth produce polarization-dependent VUV signals. Temperature scans reveal distinct maxima among specimens cut from one boule and among positions within a specimen. The source sustains operation over 33.3~h. In a separate 3-h record acquired with 1-ms gates, local referencing achieves near-Poisson averaging, reaching a fractional Allan deviation of $3.5\times10^{-5}$ at 1000~s. These results establish a practical VUV source for stable, long-duration spectroscopy of weak nuclear absorption.

\end{minipage}
\end{center}
\vspace{6pt}
\end{@twocolumnfalse}]

\begin{figure*}[b]
\centering
\includegraphics[width=\textwidth]{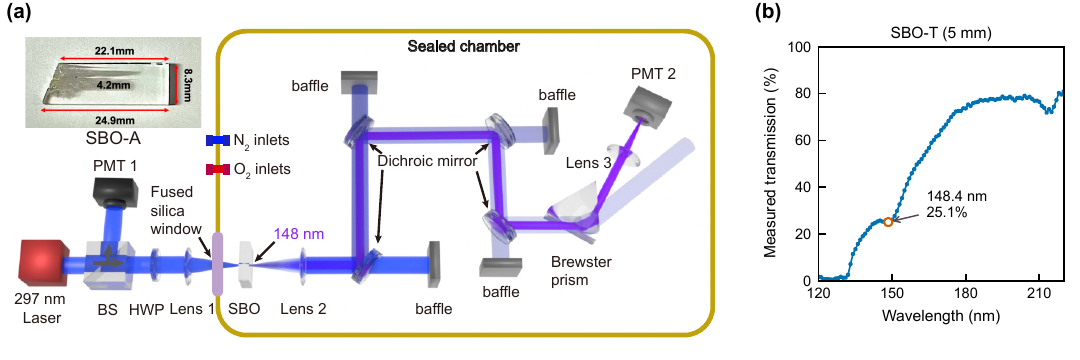}
\caption{Experimental setup and SBO transmission. (a) Final frequency-doubling stage and detection path, with a photograph of \mbox{SBO-A} inset. PMT~1 detects a beam-splitter pick-off of the 297-nm pump before the fused-silica entrance window; PMT~2 detects the separated VUV light. The sealed enclosure contains 6N nitrogen (99.9999\%) at slightly above atmospheric pressure; the surrounding optics operate in air. Gas inlet valves admit nitrogen for operation and oxygen for background measurements. In the photograph, light propagates into the page over approximately 4.2~mm. (b) Measured transmission of the 5-mm-thick specimen \mbox{SBO-T}, cut from the same boule as \mbox{SBO-A} and \mbox{SBO-B}, recorded with an ML~6500 VUV/UV system (Laser Zentrum Hannover). The measured values include surface reflection and bulk attenuation.}
\label{fig:setup}
\end{figure*}

\begin{figure*}[t]
\centering
\includegraphics[width=0.96\textwidth]{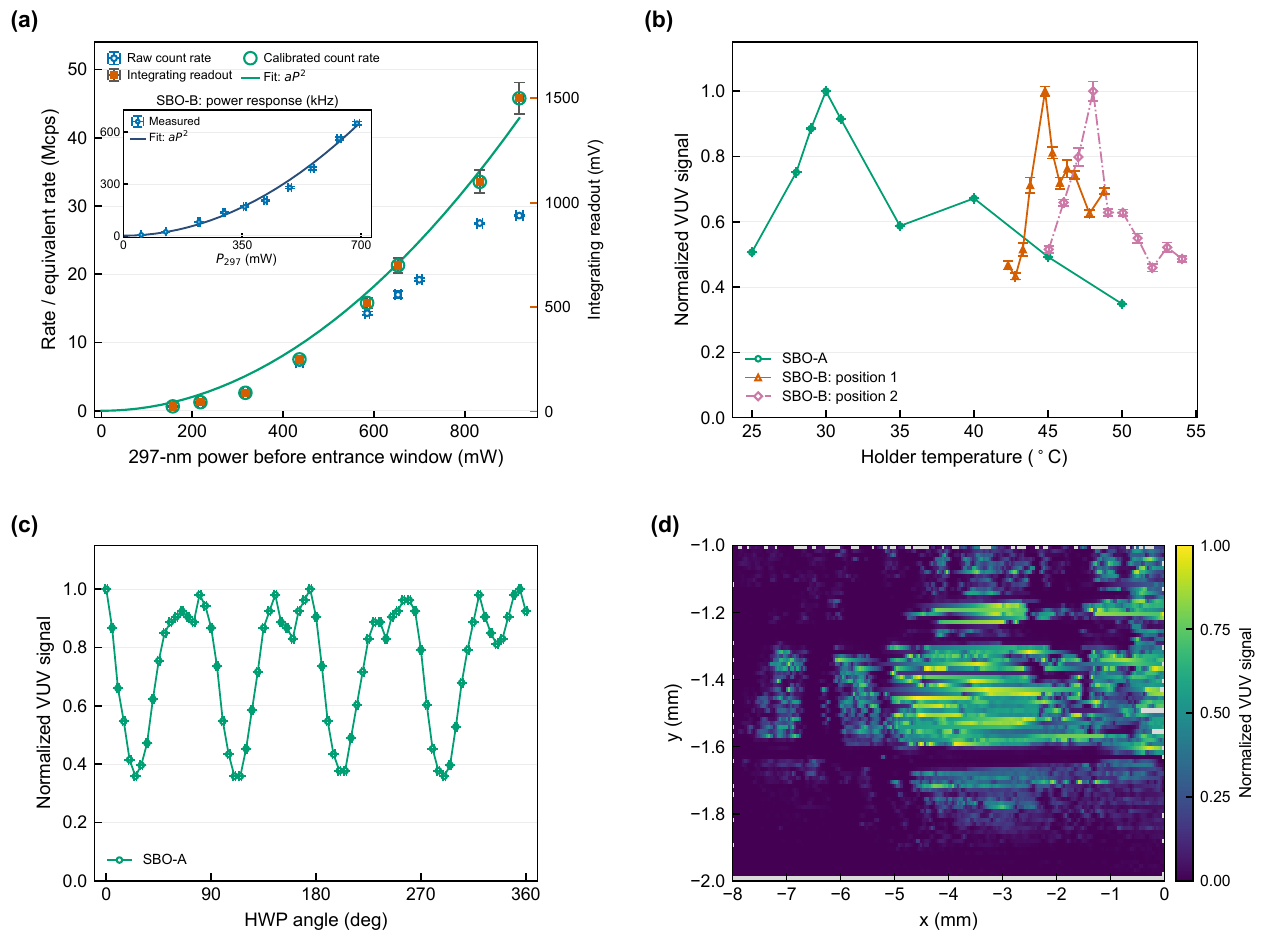}
\caption{Power, temperature, polarization, and spatial response. (a) Raw PMT count rate, current-integrating readout, and calibrated count rate for \mbox{SBO-A}. Only the calibrated rate is fitted to $aP^2$, with zero intercept. The integrating and calibrated markers share one set of error bars because they derive from the same voltage readings. Inset: \mbox{SBO-B} count-rate response fitted to the same form. Pump power is measured immediately before the fused-silica entrance window. (b) Temperature scans of two specimens cut from different regions of one boule, including two positions on \mbox{SBO-B}. Pump power is matched; focusing and polarization are fixed when comparing positions within each specimen. (c) \mbox{SBO-A} response to half-wave-plate rotation. (d) \mbox{SBO-A} spatial response at fixed holder temperature and pump polarization. In (b--d), registered PMT count rates are normalized to each curve's or map's measured maximum. Vertical bars in (b,c) are Poisson estimates for 1-s readings of \mbox{SBO-A} and standard deviations of ten readings for \mbox{SBO-B}. Other error bars are described in Sec.~\ref{sec:response}. Lines in (b,c) connect the data points to guide the eye.}
\label{fig:characterization}
\end{figure*}

\begin{figure*}[t]
\centering
\includegraphics[width=0.96\textwidth]{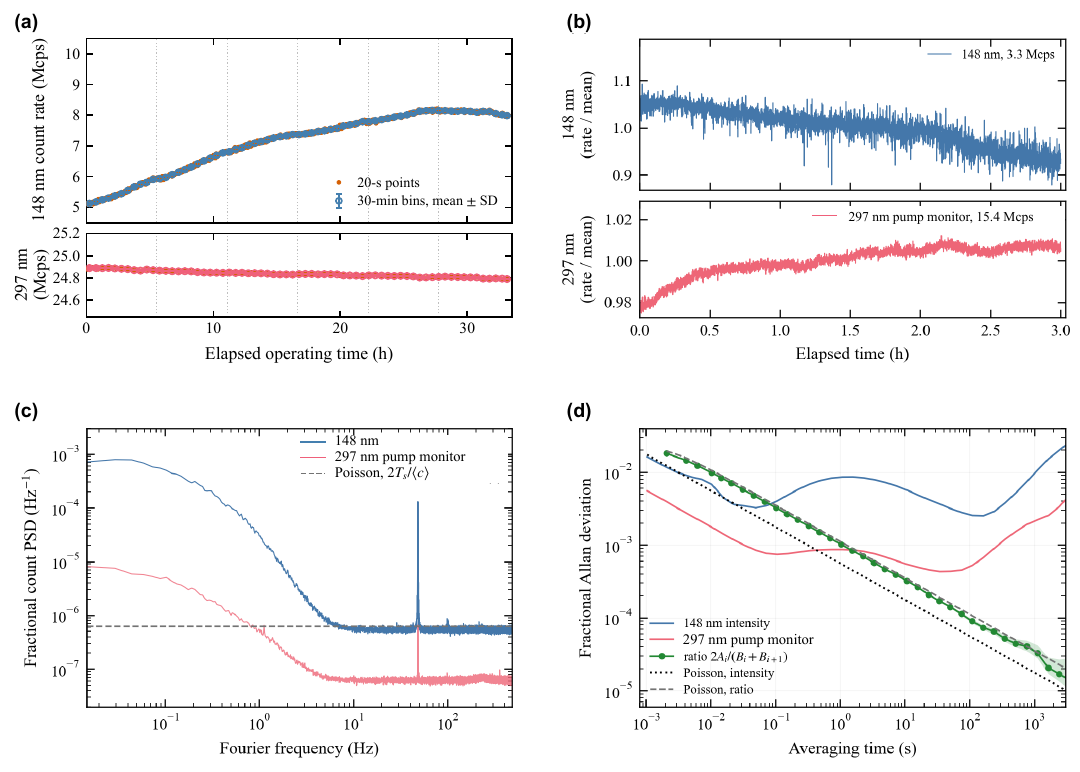}
\caption{Long-duration operation with \mbox{SBO-A} (a) and intensity noise of \mbox{SBO-B} (b--d), shown without count-rate correction. (a) Count rates over 33.3~h in nitrogen at the optimized holder temperature. Orange points are individual 20-s readings; open circles are 30-min-bin means, with bars showing the standard deviation (SD) of the readings within each bin. Dotted lines separate acquisition runs. (b) Separate 3-h record, shown in 1-s bins and normalized to each detector's mean. The gate width is $T_g=1$~ms and the measured mean period is $T_s=1.0376$~ms. (c) One-sided power spectral densities (PSDs) of fractional counts; the Poisson reference applies to the 148-nm channel. Above approximately 4~Hz, the 148-nm white floor lies 13\% below the registered-count Poisson level $2T_s/\langle c\rangle$, with $\langle c\rangle=3309.82$ counts per gate. A narrow feature near 50~Hz may originate from mains pickup. (d) Overlapping fractional Allan deviations of the VUV intensity, pump monitor, and adjacent-reference VUV ratio. The ratio lies approximately 7\% below its registered-count Poisson reference on average from 2~ms to 10~s, remaining within 8\% over this interval and within 17\% up to 1000~s. The reference includes correlations from shared reference gates; the green band is an approximate 68\% confidence interval for the measured ratio Allan deviation under a white-noise model. Estimators are given in Sec.~\ref{sec:stability}.}
\label{fig:stability}
\end{figure*}

\section{Introduction}
Optical clocks provide precise frequency references for metrology and tests of fundamental physics~\cite{Ludlow2015}. The low-lying isomer of \Th{} extends laser interrogation to a nuclear transition~\cite{Peik2003}, with enhanced sensitivity to variations of fundamental constants~\cite{Flambaum2006,Beeks2025}. Its approximately 8.4-eV excitation energy requires vacuum-ultraviolet (VUV) light near 148.4~nm. Embedding thorium in a transparent crystal provides a large absorber ensemble and a route to a solid-state nuclear clock~\cite{Rellergert2010,Kazakov2012}.

Nuclear fluorescence~\cite{Kraemer2023}, direct laser excitation~\cite{Tiedau2024,Elwell2024}, and frequency-comb spectroscopy~\cite{Zhang2024} established optical access to this transition. Continuous-wave VUV sources~\cite{Lal2025,Xiao2026} have since enabled absorption spectroscopy~\cite{Morawetz2026,Wang2026} and nuclear-clock operation~\cite{Wang2026,Toscani2026,Huang2026}. Weak-absorption readout requires sustained photon flux, rejection of intensity drift, and photon-counting sensitivity.

Strontium tetraborate (SBO) combines VUV transparency with second-order optical nonlinearity~\cite{Pan2002,Petrov2004,Khamaganova2026,Li2026ThSBO}. Random quasi-phase matching supports frequency conversion in this material~\cite{Aleksandrovsky2008,Trabs2016}, including continuous-wave generation at 148.4~nm~\cite{Lal2025}. Other crystal-based routes include cavity-enhanced continuous-wave generation in periodically poled BaMgF$_4$~\cite{Subramanian2026} and VUV frequency-comb generation in engineered quasi-phase-matched crystals~\cite{Vasilyev2026}. Here we characterize a representative SBO source, compare temperature responses across samples and positions, and assess long-duration operation and local referencing for weak-absorption readout.

\section{Experimental setup}
Three successive doubling stages convert a continuous-wave 1187-nm laser to 148.4~nm through 594 and 297~nm. The final stage uses single-pass conversion in SBO [Fig.~\ref{fig:setup}(a)], with a pump waist diameter of approximately $10~\mu$m. A half-wave plate before the sealed enclosure controls the input polarization, and a beam splitter directs a small fraction of the pump to a monitor photomultiplier tube (PMT). Pump powers are measured immediately before the fused-silica entrance window. A temperature-controlled mount and translation stage set the holder temperature and illuminated position. All reported temperatures refer to the holder.

The enclosure in Fig.~\ref{fig:setup}(a) is evacuated to approximately $10^{-5}$~mbar with a turbomolecular pump. Pumping is then stopped, and the enclosure is backfilled with 6N nitrogen (99.9999\%) to slightly above atmospheric pressure. The external optics operate in air. Nitrogen suppresses absorption by ambient oxygen and provides heat exchange for temperature control~\cite{Lal2025}. Valved inlets admit nitrogen or oxygen. Introducing a small amount of oxygen immediately extinguishes the detected 148.4-nm signal; the residual count rate defines the background. Comparisons between inert-gas and air environments have also been used to isolate VUV signals~\cite{Lal2025,Vasilyev2026}.

All crystals were grown by top-seeded solution growth (TSSG). Crystals from four sources produce continuous-wave VUV output, with differing powers. The representative specimens \mbox{SBO-A} and \mbox{SBO-B}, together with the 5-mm-thick transmission specimen \mbox{SBO-T}, were cut from different regions of a single TSSG-grown boule, then oriented and polished. The photograph in Fig.~\ref{fig:setup}(a) shows \mbox{SBO-A}; light propagates into the page over approximately 4.2~mm. Transmission of \mbox{SBO-T}, measured with an ML~6500 VUV/UV system (Laser Zentrum Hannover), is 25.1\% at 148.4~nm [Fig.~\ref{fig:setup}(b)], including surface reflection and bulk attenuation.

A Hamamatsu R8487 PMT detects the VUV light in photon-counting and current-integrating modes. The overall detection efficiency, including optical transmission and detector quantum efficiency, is $(0.756\pm0.109)\%$. We use this efficiency and the current-integration calibration to estimate the VUV power at the SBO exit. The quoted power uncertainty is propagated from the detection-efficiency budget.

\section{Output and sample-dependent response}
\label{sec:response}
Figure~\ref{fig:characterization}(a) compares photon counting with the current-integrating readout. Calibration against low-rate photon counting converts the integrating signal into an equivalent count rate and corrects saturation of the counting response. The calibrated rate and scaled integrating signal coincide by construction and share the same voltage data. Fits to $q=aP^2$ give $a=50.6$ and $1.35$~Mcps/W$^2$ for \mbox{SBO-A} and \mbox{SBO-B}, with root-mean-square residuals of 1.79 and 0.022~Mcps, respectively. The \mbox{SBO-A} fit is unweighted; \mbox{SBO-B} uses iterative effective-variance weighting with measured count-rate standard deviations and 1\% pump-power uncertainties. The exponent is fixed at two, so these fits describe the overall trend. For \mbox{SBO-B}, some mean rates differ from the fit by more than the standard deviation of ten readings at the same pump setting.

At the maximum recorded pump power of 920~mW before the entrance window, the raw rate is 28.6~Mcps (million counts per second). The integrating readout gives a calibrated rate of 45.8~Mcps and an estimated SBO output of $(8.1\pm1.2)$~nW. This nominal output is about six times the $1.3^{+0.7}_{-0.6}$~nW reported by Lal \textit{et al.} at 325~mW incident pump power~\cite{Lal2025}. The measurements differ in pump power, focusing, and the locations at which power is specified; here the pump is measured before the entrance window and the VUV output is estimated at the SBO exit. Dividing the nominal outputs by the reported pump powers squared gives approximately 9.6 and 12~nW/W$^2$, respectively. The higher output therefore does not alone imply a larger conversion coefficient.

At a pump power of 920~mW, oxygen introduced into the nitrogen-filled enclosure absorbs the VUV light, leaving a background rate of 1.22~kHz. The registered VUV rate before oxygen admission gives a signal-to-background ratio of $2.3\times10^4$. Using the calibrated rate of 45.8~Mcps gives an estimated ratio of $3.8\times10^4$. These ratios characterize the maximum-power measurement; they are distinct from the long-duration record below.

Temperature scans at matched pump power show maxima near $30^\circ$C for \mbox{SBO-A} and near $45^\circ$C and $48^\circ$C at two positions on \mbox{SBO-B} [Fig.~\ref{fig:characterization}(b)]. Focusing and polarization remain fixed when comparing positions within each specimen. The sampled maxima differ by about $3^\circ$C within \mbox{SBO-B} and by $15$--$18^\circ$C between \mbox{SBO-A} and \mbox{SBO-B}, although both were cut from one boule. This variation motivates local temperature optimization.

Polarization-dependent VUV signals are observed in multiple crystals. For \mbox{SBO-A}, the half-wave-plate scan gives a maximum-to-minimum registered count-rate ratio of approximately 2.8 [Fig.~\ref{fig:characterization}(c)]. At fixed holder temperature and pump polarization, the spatial scan reveals a nonuniform VUV response [Fig.~\ref{fig:characterization}(d)].

The calibration $q_{\rm cal}=f(q_{\rm reg})$ maps the registered PMT rate $q_{\rm reg}$ to its calibrated equivalent $q_{\rm cal}$ by smooth interpolation of the two-mode comparison. Cross-checks at different integrating gains under otherwise unchanged conditions support readout linearity, with no observed clipping at 1.5~V.

For \mbox{SBO-A} in Fig.~\ref{fig:characterization}(a), raw-rate and pump-power bars show the recorded half-ranges: $28.6\pm0.1$~Mcps at $920\pm8$~mW, with an integrating readout of 1500~mV. The approximately $\pm5\%$ spread in duplicate voltage readings is propagated to the calibrated rate with the calibration held fixed. The shared bars exclude calibration uncertainty. \mbox{SBO-B} vertical bars are sample standard deviations of ten readings, with 1\% pump-power uncertainties. For \mbox{SBO-A} temperature and polarization scans, the 1-s acquisition time ($T=1$~s) gives Poisson estimates $\sigma_q=\sqrt{q/T}$ rather than measured repeatability. The 10-mK horizontal temperature bars represent the controller\textquotesingle s specified stability. Horizontal angular bars are $1^\circ$. Normalized values and bars use a fixed measured maximum and exclude its uncertainty.

\section{Long-duration operation and intensity noise}
\label{sec:stability}
Figure~\ref{fig:stability} shows registered count rates without correction. The 33.3-h record [Fig.~\ref{fig:stability}(a)] was acquired with \mbox{SBO-A} in nitrogen at its optimized holder temperature during nuclear absorption spectroscopy~\cite{Wang2026}. Six acquisition runs provide 1032 measurements, each integrating 20~s on both the VUV and pump-monitor detectors. The plotted readings provide 5.73~h of accumulated integration per detector, approximately 17\% of the 33.3-h operating period. This intermittent sampling tracks long-duration operation; 30-min binning highlights the slow evolution. The VUV rate rises from approximately 5 to 8~Mcps while the pump monitor remains near 25~Mcps, demonstrating sustained operation despite slow drift. Changes in frequency-doubling efficiency, optical alignment, path transmission, or detector response may contribute; their individual contributions are unresolved.

A separate 3-h record acquired with \mbox{SBO-B} using 1-ms gates has mean detected rates of 3.3~Mcps for the VUV signal and 15.4~Mcps for the pump monitor [Fig.~\ref{fig:stability}(b)]. The \mbox{SBO-B} power scan, temperature scans, and stability record were acquired separately; their absolute count rates are not compared at a common operating point. The VUV rate decreases slowly while the pump monitor drifts upward. The pump-power servo, SBO temperature control, and laser frequency lock remain active without loss of lock. Above approximately 4~Hz, the fractional count-noise power spectral density (PSD) approaches a white floor [Fig.~\ref{fig:stability}(c)].

For local normalization, we divide the VUV gate counts $c_i$ into alternating signal and reference sequences, $A_i=c_{2i+1}$ and $B_i=c_{2i}$, and form
\begin{equation}
R_i=\frac{2A_i}{B_i+B_{i+1}}.
\label{eq:ratio}
\end{equation}
This offline construction uses adjacent VUV gates as local references. Its overlapping fractional Allan deviation~\cite{Riley2008} decreases approximately as $\tau^{-1/2}$ into the $10^3$-s range [Fig.~\ref{fig:stability}(d)], reaching $3.5\times10^{-5}$ at 1000~s by log--log interpolation between neighboring points. A log--log fit over 100--1000~s gives a slope of $-0.438\pm0.083$, consistent with white-noise averaging. The $1\sigma$ uncertainty is the larger of the formal fit standard error and the standard deviation from white-noise Monte Carlo trials.

The noise analysis uses 10,412,032 gates of width $T_g=1$~ms and measured mean period $T_s=1.037554$~ms, giving a duty factor $d=T_g/T_s=0.9638$. The mean VUV count is $\mu=\langle c\rangle=3309.82$ per gate, equivalent to $\bar q=\mu/T_g=3.31$~Mcps during the gate. The acquisition contains short inter-block gaps; the PSD and Allan estimators approximate this nonuniform sampling by the mean period. No gates are rejected or globally detrended.

\needspace{8\baselineskip}
We estimate one-sided PSDs of fractional counts $y_i=c_i/\mu-1$ by Welch's method~\cite{Welch1967}, using 65,536-gate Hann windows with 50\% overlap and subtraction of each segment mean. The ideal Poisson level is
\begin{equation}
S_{\rm P}=\frac{2T_s}{\mu}=\frac{2}{\bar q d}.
\end{equation}
The VUV white floor, evaluated as the median PSD over 20--400~Hz, is $5.45\times10^{-7}$~Hz$^{-1}$, 13\% below the Poisson level of the registered counts. A narrow technical-noise feature near 50~Hz may originate from mains pickup. The 297-nm spectrum characterizes the registered pump-monitor counts; no independently calibrated photon-noise limit is assigned to that channel.

Both intensity and ratio series are normalized to their means before calculating the Allan deviation. For $m$ ratio samples and $\tau=2mT_s$, the Poisson reference is $\sigma_{R,\mathrm{P}}^2=(2m-3/4)/(m^2\mu)$, including correlations from shared reference gates. The intensity reference is $\sigma_{I,\mathrm{P}}^2=T_s/(\mu\tau)$. The measured ratio deviation is approximately 7\% below its reference on average from 2~ms to 10~s, remaining within 8\% over this interval and within 17\% up to 1000~s. Its offset agrees in magnitude with the 13\% lower PSD floor, since Allan deviation scales with the square root of white-noise PSD. These comparisons use registered counts and do not establish an independently calibrated photon-noise limit. The green band gives an approximate 68\% confidence interval for the measured ratio Allan deviation using a white-noise $\chi^2$ model~\cite{Riley2008}.

Local referencing suppresses the slow drift that limits averaging of the VUV intensity. The resulting near-Poisson averaging supports differential readout of weak absorption. In an ideal shot-noise-limited measurement, fractional sensitivity scales as $(q_{\rm det}\tau)^{-1/2}$, where $q_{\rm det}$ is the detected photon rate. Higher VUV power can therefore improve sensitivity if detector response and drift rejection are maintained. This offline test establishes the averaging performance of intensity referencing, rather than a measured absorption detection limit.

\section{Conclusion}
We demonstrate single-pass continuous-wave generation at 148.4~nm in TSSG-grown SBO, reaching an estimated output of 8.1~nW at 920~mW pump power. At this power, the calibrated signal-to-background ratio is approximately $3.8\times10^4$. Separate records demonstrate operation over 33.3~h with \mbox{SBO-A} and near-Poisson averaging with \mbox{SBO-B}, reaching $3.5\times10^{-5}$ at 1000~s. Polarization and temperature scans guide local optimization, while the 25.1\% transmission of 5-mm-thick \mbox{SBO-T} at 148.4~nm supports VUV extraction through millimeter-scale SBO. These results support sustained differential measurements of weak nuclear absorption, with further sensitivity gains expected from higher detected photon flux.

Domain control offers a route to improved conversion. Li doping modifies domain formation in SBO and has enabled 266-nm frequency doubling~\cite{Wu2024Li}. Together with VUV generation in engineered quasi-phase-matched crystals~\cite{Vasilyev2026}, this motivates exploring domain control for SBO while assessing the accompanying pump and VUV losses.

Cavity-enhanced generation at 148.4~nm has been demonstrated in periodically poled BaMgF$_4$~\cite{Subramanian2026}. The 5-mm-thick \mbox{SBO-T} transmits 81\% at 220~nm; higher transmission at 297~nm is plausible beyond the reported absorption band near 210~nm~\cite{Pan2002,Yang2026}. For a 1-mm SBO crystal, we assume 98\% net mode-preserving single-pass transmission at 297~nm, including surface losses at the operating angle and polarization. Three 99\%-reflecting mirrors and matched input coupling then give an ideal bow-tie power buildup of approximately 20. If the 1-mm crystal retains the conversion coefficient measured over the 4.2-mm optical path at the same focus, quadratic scaling from 8.1~nW predicts approximately $3.4~\mu$W. This conditional projection requires low crystal and cavity losses, stable locking, and sustained conversion at higher circulating power.

\section*{Funding}
This work is supported by the National Natural Science Foundation of China (Grant No.~92476001), the Chinese Academy of Sciences (CAS) Project for Young Scientists in Basic Research (Grant No.~\mbox{YSBR-085}), the Anhui Initiative in Quantum Information Technologies (Grant No.~AHY040200), the Shanghai Municipal Science and Technology Major Project (Grant No.~2019SHZDZX01), the Independent Deployment Project of Hefei National Laboratory (Grant No.~ZB2025010100), the Quantum Science and Technology--National Science and Technology Major Project (Grant Nos.~2021ZD0300106, 2021ZD0302002, and 2021ZD0300902), the Strategic Priority Research Program of CAS (Grant Nos.~XDA0520101 and XDB0920200), and the New Cornerstone Science Foundation.

\begin{acknowledgments}
We thank Jian-Wei Pan and Yu-Ao Chen of the University of Science and Technology of China for their guidance in designing the experiment and for helpful discussions and suggestions, and their colleague Guang-Yue Hu for assistance in checking the optical setup and advice on VUV optics. We thank Linqiang Hua of the Innovation Academy for Precision Measurement Science and Technology, Chinese Academy of Sciences, for advice on VUV detection and for providing optical components for 148-nm light. We also thank Kemi Xu and Ke Zhang of Beijing Institute of Technology for helpful comments on the manuscript.
\end{acknowledgments}

\section*{Disclosures}
Authors affiliated with Shanghai Precilasers Technology Co., Ltd. are employees or owners of the company. The authors declare no other conflicts of interest.

\section*{Data availability}
The data and code supporting this study are available from the authors upon reasonable request.

\end{document}